\documentclass[12pt]{article}

\pdfoutput=1
\usepackage{graphicx}
\usepackage{graphics}
\usepackage{multirow}

\title{First experience in operating the population of the condition databases for the CMS experiment}

\author{ M.De Gruttola$^{1,2,3}$, S.Di Guida$^{1}$, D.Futyan$^{4}$, F.Glege$^{2}$, G.Govi$^{5}$, \\  
V.Innocente$^{1}$, P.Paolucci$^{2}$, A.Pierro$^{8}$, D.Schlatter$^{1}$\\
\\  
\small $^1$CERN, Geneva, Switzerland,   \\       
\small$^2$INFN Sezione di Napoli, Naples, Italy, \\
\small$^3$Universit\`a  degli studi di Napoli ``Federico II'', Naples, Italy, \\
\small $^4$Imperial College, London, UK,\\
\small $^5$Department of Physics, Northeastern University, Boston, MA. 02115, USA,\\
\small $^6$Universit\`a di Milano - Bicocca, Dipartimento di Fisica G. Occhialini, Milan, Italy,\\
\small$^7$Centre de Calcul de l'IN2P3, CNRS, Lyon, France,\\
\small$^8$INFN Sezione di Bari, Bari, Italy, \\
\small$^9$Department of Physics, Princeton University, Princeton, NJ.08542, USA.\\
\\
email:michele.de.gruttola@cern.ch
}
\begin{document}

\maketitle                                    

\begin{abstract}
Reliable population of the condition databases is critical for the correct operation of the online selection as well as of the offline reconstruction and analysis of data. 
We will describe here the system put in place in the CMS experiment to populate the database and make condition data promptly available both online for the high-level trigger and offline for reconstruction. 
The system, designed for high flexibility to cope with very different data sources, uses POOL-ORA technology in order to store data in an object format that best matches the object oriented paradigm for \texttt{C++} programming language used in the CMS offline software. 
In order to ensure consistency among the various subdetectors, a dedicated package, PopCon (Populator of Condition Objects), is used to store data online. 
The data are then automatically streamed to the offline database hence immediately accessible offline worldwide. 
This mechanism was intensively used during 2008 in the test-runs with cosmic rays. 
The experience of this first months of operation will be discussed in detail. 
\end{abstract}

\section{Introduction}
Databases have become a vital part of the LHC experiments's software. 
The large amount of data needed to describe, set up and operate the detectors makes the DB system an essential service for the experiments to run: these data, indeed, are used for the calibration of the physical responses of the detectors themselves. 
Therefore, a project was started inside CMS many years ago to set up a system able to populate the condition data efficiently. 

This system has been successfully deployed and operated nicely last year during cosmic runs with and without magnetic field.    
Important requirements constraining the possible DB model design are:
\begin{itemize}
\item CMS always requires to be able to operate without network connection to the outside world (IT Meyrin included). Therefore, an independent DB infrastructure should reside in CMS P5 network.
\item Condition/calibration data access for the offline reconstruction shall hide to the user any underlying database technology. The storing and access mechanism should best match the object oriented paradigm for the \texttt{C++} programming language used in the CMS offline software (\texttt{CMSSW}\cite{CMSSW})  
\item Offline condition data work-flow should fit a multi-tier structure as in the case for the event data.
\end{itemize}

Having to tackle all these constraints, the team involved in the CMS DB project\cite{twikiDB}, working in collaboration with the CERN-IT department, chose to rely on 3 database instances for storing non-event data:
\begin{enumerate}
\item {\textbf{OMDS}}  ({\textbf{O}}nline {\textbf{M}}aster {\textbf{D}atabase} {\textbf{S}}ystem) is located in the online network at IP5; it stores the data needed for the configuration and proper settings of the detector, and the condition data produced directly from the front-end electronics. 
All tables contained in it are purely relational.
\item {\textbf{ORCON}} ({\textbf{O}}ffline {\textbf{R}}econstruction  {\textbf{C}ondition} DB {\textbf{ON}}line subset) is also located in the online  network: it stores all the condition data needed for the reconstruction  of physics quantities as well as for detector performance studies. 
These are a small subset of all the online quantities.
The data in it are written using the POOL-ORA\cite{POOLORA} technology and are retrieved by the HLT programs as {\texttt{C++}} objects for the offline software.  
\item  {\textbf{ORCOFF}} ({\textbf{O}}ffline {\textbf{R}}econstruction  {\textbf{C}ondition} DB {\textbf{O}}ffline subset) is located at the Tier-0 site (CERN Meyrin): it contains a copy of ORCON, made through ORACLE streaming. It contains the entire history of all CMS condition data and serves as an input source for both prompt reconstruction and the condition deployment service at Tier-1/Tier-2 sites. 
Data contained in it are retrieved by the reconstruction algorithms as \texttt{C++} objects for the offline software. 
\end{enumerate}

The actual policy of the CMS community is to write any condition/calibration data needed for offline purposes in ORCON. ORACLE streaming provides the transfer from ORCON to ORCOFF. 
In this paper we will mainly describe the condition data work-flow, i.e. non event data, stored in ORCON/ORCOFF. Attention will be focused on the system set-up to populate centrally the CMS condition database and to monitor the database activity itself. 

\section{Condition data description}
Non-event data can be classified in three main groups:
\begin {itemize}
\item {\bf Configuration data}: the data needed to bring the detector in running mode. 
This class includes voltage settings of power supplies, gas pressures, and any programmable parameter for front end electronics and trigger. 
\item {\bf Condition data}: the data from any detector subsystem describing its state, usually uploaded in OMDS directly from detector back-end devices. For example the Data Control System (DCS) information are stored with the ORACLE interface provided by PVSS\cite{PVSS}. 
Only a subset of these data is transferred to the offline system for detector performance studies.  
\item {\bf Calibration data}: the data describing the calibration and alignment of the single pieces of the different sub-detectors. 
These quantities (such as pedestal offsets, drift values, noise, alignments, etc) are evaluated by running offline dedicated algorithms. 
Since these data are needed both online, in order to be used by the HLT algorithms, and offline, in order to reconstruct properly the physical quantities coming from collision events, they must be stored in the offline condition databases.
Therefore, they should match the corresponding raw data coming from the collision events revealed by the detector.
\end{itemize}   

All these data need a tag and an interval of validity as meta-data. The interval of validity (IOV) is the contiguous (in time) set of events for which non-event data are to be used in reconstruction.
According to the use-case, the IOV will be defined in terms either of GPS-time (mainly for condition data) or run-number\footnote[5]{Progressive number given to a set of contiguous Physics events, coming from either collisions of particles accelerated in the LHC or cosmic rays revealed by the detector.} range (usually for calibrations). 
While the IOV for some electronic related conditions (pedestals and noise) is identical to the time interval in which these data were used in the online operations, some calibration data may posses an IOV different from the time range in which they have been calculated.
For this reason, the IOV assignment for a given set of condition data is carried out at the offline level.
Each payload object, i.e. each data stored as POOL-ORA object in ORCON/ORCOFF, is indexed by its IOV and a  tag, a label describing the calibration version, while the data themselves do not contain any time validity information; when new better calibrations are evaluated, the tag labelling the data should be changed.     

The matching with the raw data from the collision events is indeed possible via these meta-data: the reconstruction algorithms for the analysis of a given run query the offline condition data corresponding to the same run grouped through a set of tags, called \emph{global tag}.

\section{Database architecture}
Different data usage and access from online to offline determines the overall architecture.
In the online network, the data are mainly written into the database. 
Data size is expected to become very large (several TBs), and, since condition data will constantly flow into the DB, the time needed to store these data in OMDS is a critical issue. 
The online data are stored at random time, and the time when their storage occurs is usually not synchronous with respect to the time when they are read, since these data can be taken by different sources. 
Furthermore, different data items must be accessible in order to be compared between each other. 
Therefore, OMDS is designed with relational schemas: each sub-detector group designed its own DB schema, reflecting as much as possible the detector structure.

On the contrary, in the offline network data are mainly read from the databases.  These data must be synchronized with the event reprocessing, and grouped before they are read, so that they can be decoded according to predefined rules. An object oriented solution has been adopted for data stored in ORCON/ORCOFF.

\begin{figure}[hbtp]
  \begin{center}
    \resizebox{1.0\textwidth}{!}{\includegraphics{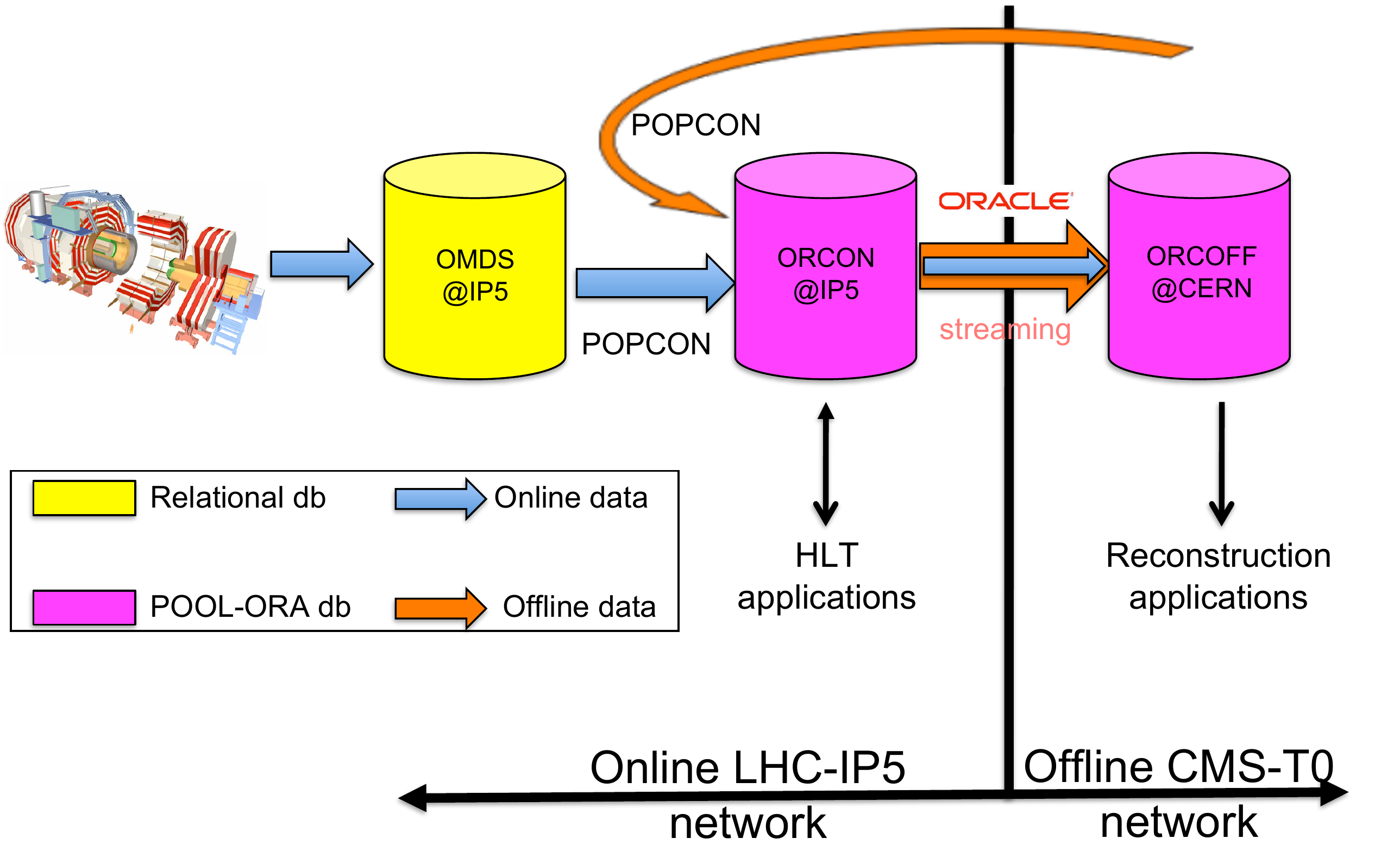}} \caption{CMS condition databases architecture.}
    \label{fig:CondDBArchitecture}
  \end{center}
\end{figure}

The general data flow of non event data is the following\cite{DBCHEP}: configuration data are prepared using the equipment management information and are loaded into the detector (hardware and software). 
During data taking, the detector produces condition data, which are first stored in OMDS. 
The offline conditions subset is extracted and sent to the offline sites, as shown in Figure \ref{fig:CondDBArchitecture}. 
The condition data needed by the HLT farm are loaded from ORCON. 

A software application named PoPCon (Populator of Condition Objects) operates the online to offline condition data transfer and encapsulates the relational data as POOL-ORA objects. 
PopCon adds meta-data information (tag and IOV) to the condition data, so that they can be read by the offline/HLT software. 

Finally, data are transferred to ORCOFF, which is the main condition database for the CMS Tier-0, using ORACLE streaming.

From ORCOFF data will be distributed to the other tier centers, through Frontier\cite{FRONTIER} packages. 
Calibration data evaluated offline will be written to ORCON, using PopCon. 
Collision event data are therefore processed using the offline condition data. 
As data taking proceeds, we can understand better and better how the detector works; therefore, this will require new calibrations, hence new versions of condition data, identified by new tags.

\subsection{PopCon}\label{subsect:POPCON}

PopCon\cite{twikiPopCon} transfers the conditions objects from a user-defined data source to the off-line database.

Popcon is based on the cmsRun infrastructure\cite{CMSSW}, so the base PopCon application class is the EDAnalyzer\cite{EDM}. 
However, it is possible to use different data sources such as databases, ROOT files, ASCII files, etc. 
For each conditions object (payload) class a PopCon application is created.

\begin{figure}[hbtp]
  \begin{center}
    \resizebox{1.0\textwidth}{!}{\includegraphics{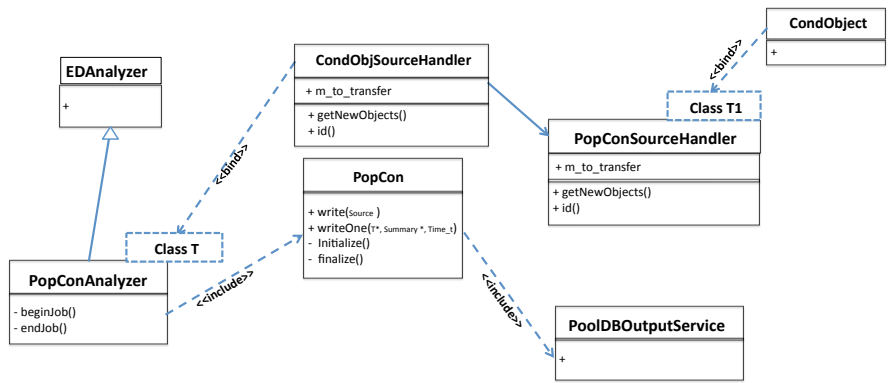}} \caption{Schema of the classes for the PopCon package.}
    \label{fig:PopConSchema}
  \end{center}
\end{figure}

The core framework consists of three parameterized classes, as can be seen in Figure \ref{fig:PopConSchema}:

\begin{itemize}
\item PopCon
\item PopConSourceHandler
\item PopConAnalyzer 
\end{itemize}

The ``detector user'' provides the code which handles the data source and specifies the destination for the data, writing a derived class of  PopConSourceHandler, where all the online (source handling) code goes. 
The user instantiates his/her objects, provides the IOV information for such objects and configures the database output module. 
PopCon configuration file associates the tag name defined according to some specific rules, to the condition object. 
Once the object is built, the PopCon application writes the data to the specified database. 
Subdetector code does not access the target output database: it only passes the objects to the output module.

The analyzer object holds the source handling object. It also serves some additional functionality such as:
\begin{itemize}
 \item Locking mechanism
 \item Transfer logging
 \item Payload verification (IOV sequence)
 \item Application state management
 \item Database output service 
\end{itemize}
The writer in PopCon iterates over the container of user objects and stores it in the user-configured data destination. 

Any transaction towards ORCON is logged by PopCon, and the process information is sent to a database account. A monitoring tool for this information was developed, in order to check the correctness of the various transactions, and to keep trace of every upload for condition data.
 
\section{First experience in operating the population of the condition DB in 2008}

In the 2008 global runs (with or without the magnetic field) the great majority of the condition data was transferred offline using a PopCon application. 
Great effort was devoted by the CMS database project team in the integration of all the software and the infrastructural chain used to upload the calibration constants into the CMS condition databases. 
Many tools were provided in order to help the sub-detector responsible people to populate the database. 
Indeed, a central procedure, based on an automatic uploader into ORCON on a dedicated machine in the online network, was successfully deployed during 2008, and will be the recommended way to populate ORCON during 2009 data taking.

\subsection{Condition objects written with PopCon in 2008}

As stated above, each piece of condition data (pedestals, Lorentz Angles, drift time, etc.) corresponds to a {\texttt C++} object (``CondObjects'') in the CMS software. 
Each object is associated with a PopCon application which writes the payload into ORCON. 
Table \ref{tab:page_layout} lists all the CondObjects used in 2008, grouped according to the subsystem they belong to. 
For each object the type, the approximate data size in ORCON and the upload frequency are also reported.
   
 \begin{table}[htbp]
    \caption{2008 CMS condition objects list}
    \label{tab:page_layout}
    \begin{center}
   \begin{tabular}{lllll} \\
            Subsystem   & Name & Type & Data size & Frequency\\  
  \multirow{3}{*} {Pixel} & SiPixelFedCablingMap & online configuration& 1K& once (before the run )  \\      
 & SiPixelLorentzAngle & offline calibration & 1MB & each run (if different) \\ 
 & SiPixelCalibConfiguration  & online calibrations & 5KB & each calibration run \\ 
  \multirow{5}{*} {Tracker} & SiStripFedCabling & online configuration& 1K&once \\        
& SiStripBadStrip & online condition & 1MB&  each run (if different)\\ 
& SiStripTreshold & offline calibration & 1MB&  each run (if different)\\        & SiStripPedestals & offline calibration & 1MB&  each run (if different)\\  
& SiStripNoise & offline calibration & 1MB&  each run (if different)            
\\ 

  \multirow{2}{*} {Ecal} & EcalPedestals & online calibration& 2MB& daily \\     & EcalLaserAPDPNRatios & online calibration & 2MB& hourly        
\\ 

 \multirow{5}{*} {Hcal} & HcalElectronicsMap & online configurations & 1MB& once (before the run)  \\  
& HcalGains & offline calibrations & 1MB& each run \\  
& HcalPedestals & offline calibrations & 1MB& each run \\  
& HcalPedestalsWidths & offline calibrations & 1MB& each run \\  
& HcalQIEData & online calibrations & 1MB& each run 

\\ 

 \multirow{7}{*} {CSC} & CSCChamberMap & online configuration & 10KB & monthly  \\  
& CSCCrateMap & online configuration & 10KB & monthly  \\  
& CSCDDUMap & online configuration & 10KB & monthly  \\  
& CSCChamberIndex & online configuration & 10KB & monthly  \\  
& CSCGains & offline calibrations & 2MB& each run  \\  
& CSCNoiseMatrix & offline calibrations & 2MB& each run  \\  
& CSCPedestals & offline calibrations & 2MB& each run  \\

\multirow{5}{*} {DT} & DtReadOut & online configuration & 10MB & once  \\  
 & DtCCBConfig &  online configuration  & 100KB & once (before the run)   \\
 & DtT0 & offline calibration & 10MB & rare  \\
 & DtTTrig & offline calibration & 1MB & at trigger change   \\
 & DtMTime & offline calibration & 1MB & daily   \\

\multirow{3}{*} {RPC} & RPCEMap & online configuration & 10MB & once  \\
& L1RPCConfig & online configuration & 10MB & once  \\
& RPCCond & online conditions & 10MB & daily  \\

\multirow{1}{*} {DAQ} & RunSummary & run conditions & 10KB & run start/end  \\

     \end{tabular}
    \end{center}
  \end{table}

\begin{figure}[h]
  \begin{center}
    \resizebox{1.0\textwidth}{!}{\includegraphics{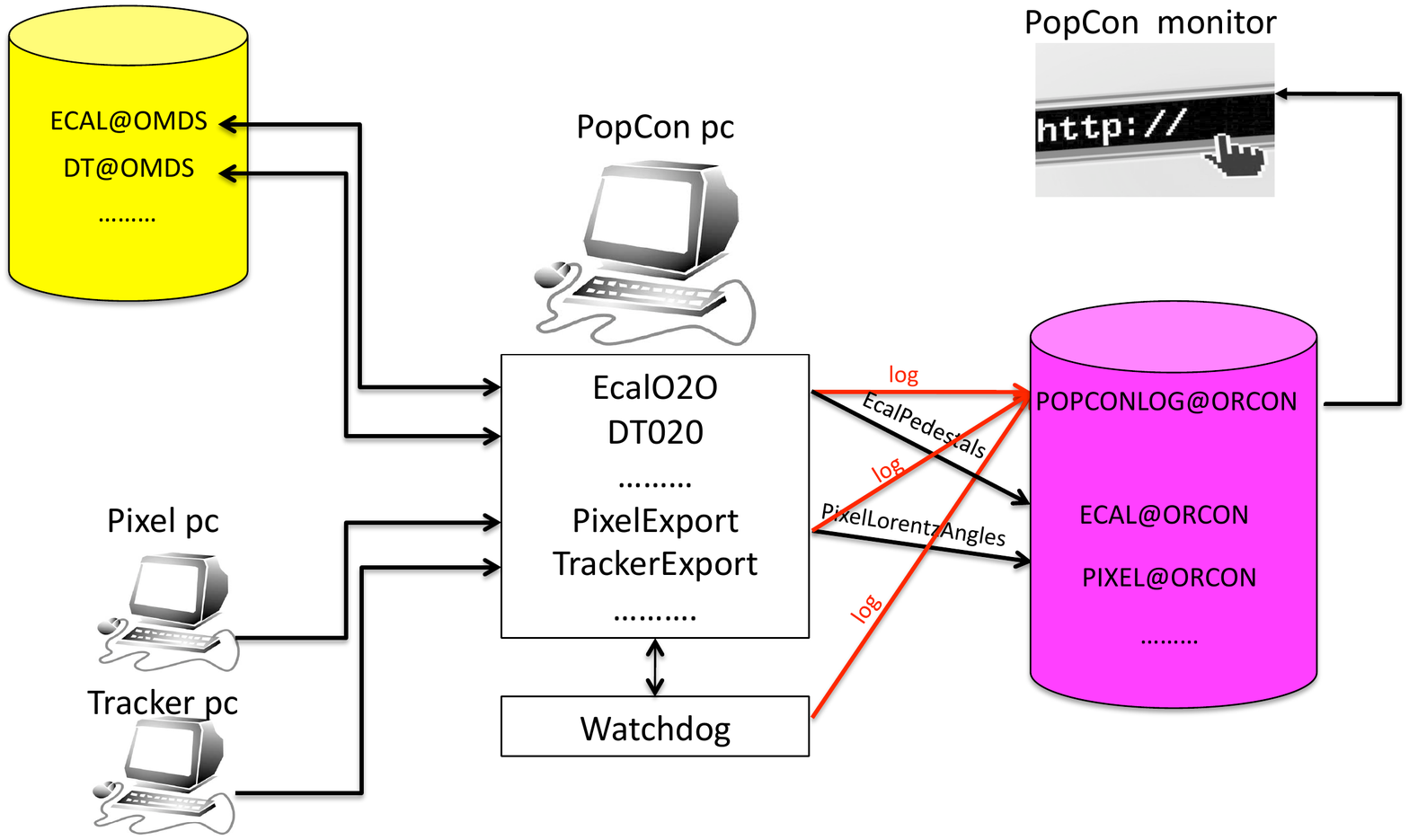}} \caption{Schematic illustration of the central system to populate ORCON, and of the web monitoring system.}\label{fig:DropBox}
  \end{center}
\end{figure}

\subsection{Central population of the condition database}

A central procedure\cite{twikiPopConOperation} was set up in 2008 for populating the CMS condition databases: it exploits a central account, explicitly devoted to condition database population, in the CMS online network. 
On that account a set of automatic jobs were centrally set up for any single sub-detector user, in order to both populate ORCON and monitor any transaction to it.

Two possibilities are given to users:
\begin{enumerate}
\item running automatically the application that reads from any online source, assigns tag and interval of validity, and uploads the constants into ORCON (mainly for condition). 
The time interval of the automatic jobs are negotiated with the users.  
\item using the account as a drop-box: users copy the calibrations in a light format into a dedicated folder, one for each sub-detector, and then these data are automatically exported in ORCON (mainly for offline calibrations).  
\end{enumerate}

Figure \ref{fig:DropBox} shows a sketch of the central system used to populate the condition database. Each sub-detector may transfer the payload onto the central PopCon PC, that then  automatically manages the exportation into the ORCON DB (using a specific set of Subdetector Exports scripts). Other automatic scripts (e.g. ECAL020, DT020 ...) check to see if new conditions have appeared in the online table, and, if so, perform the data transfer from OMDS to ORCON.   

The PopCon applications transfer each payload into the corresponding account, and all the log information in the PopConLog account on ORCON itself.

Each automatic job is associated with a ``watchdog'' tool that monitors its status.
The job monitoring information are also logged into the PopConLog account on ORCON.

A dedicated web interface, {\it PopCon monitor web interface}, was set up on a CMS web server in order to provide access to all the logged information for monitoring purposes. 
The monitor system is made of three layers:
\begin{itemize}
\item {\it Python} code to query the PopConLog account.   
\item {\it Python-JSON} code to produce a JSON (JavaScript Object Notation) string. 
\item CSS web interface to configure the look and fill of the overall information.
\end{itemize} 

\begin{figure}[h]
  \begin{center}
    \resizebox{1.0\textwidth}{!}{\includegraphics{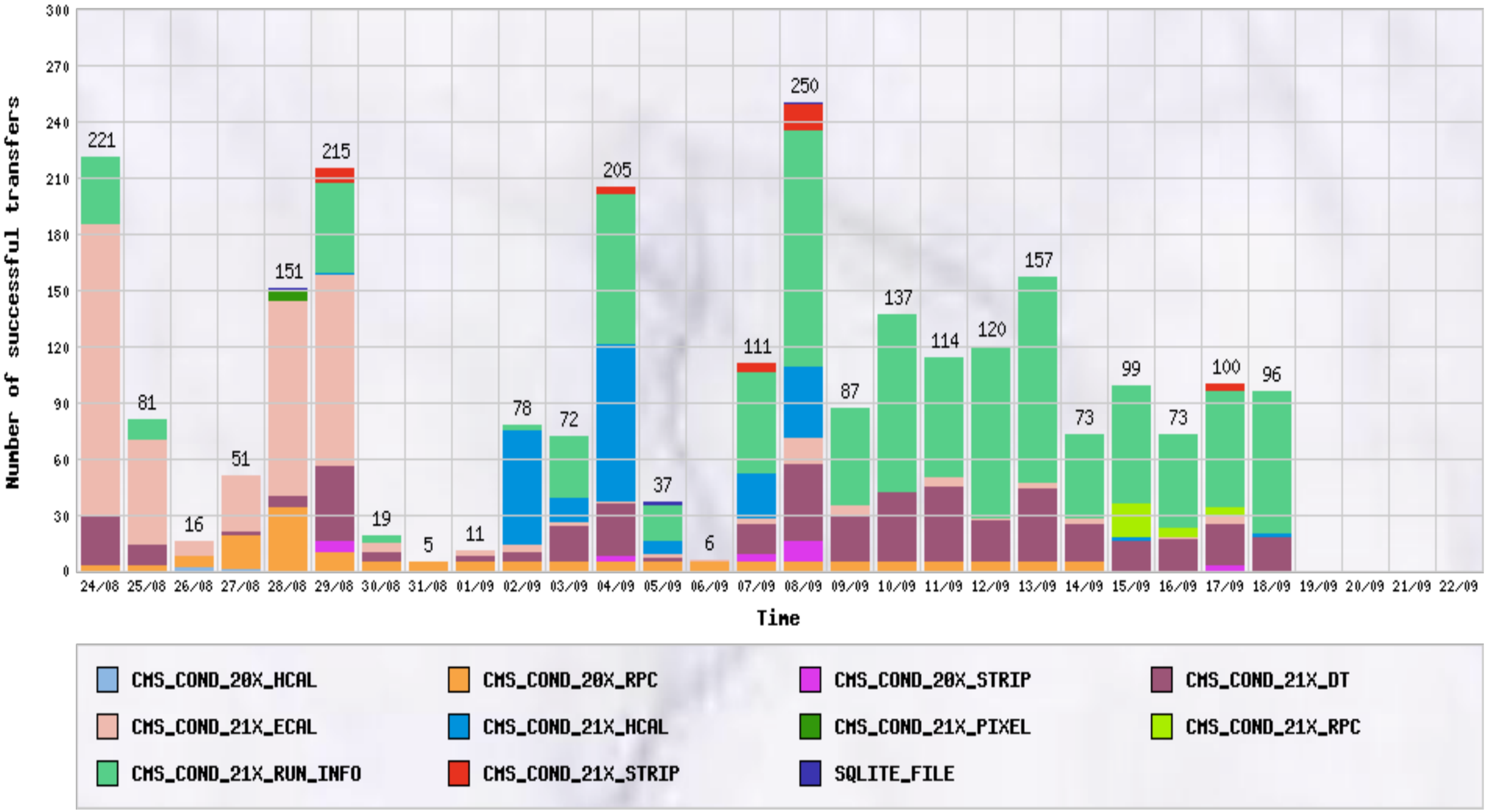}} \caption{PopCon activity between end September-beginning of October 2008.}
    \label{fig:PopConAct}
  \end{center}
\end{figure}
\begin{figure}[h]
  \begin{center}
    \resizebox{1.0\textwidth}{!}{\includegraphics{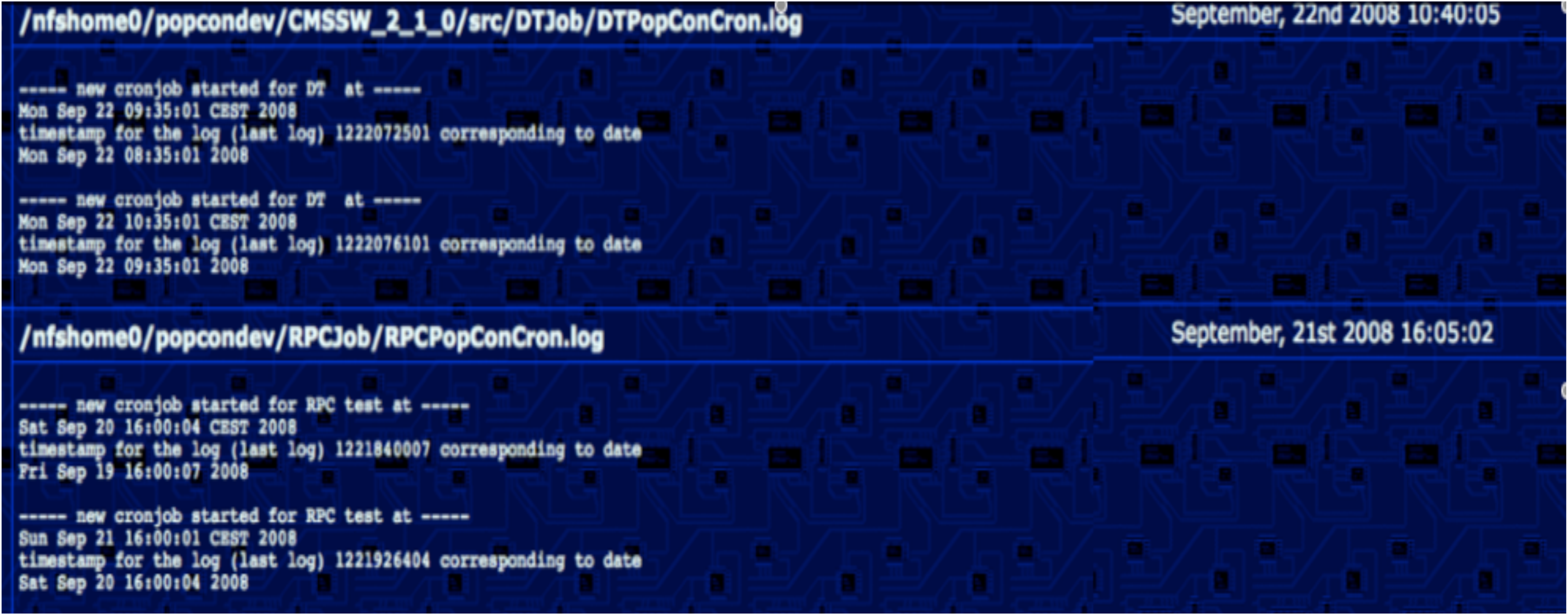}} \caption{Screenshot of the web page produced by the monitoring system that checks the watchdog tools for the automatic population of ORCON.}
    \label{fig:WatchDog}
  \end{center}
\end{figure}

Two important monitor web pages are then produced:
\begin{enumerate}
\item an activity summary, in which the number of ORCON transactions, the subsystem involved, the IOV and tag can be displayed, according to the users' requests. An example is shown in Figure \ref{fig:PopConAct}. 
\item the logs of all the central scripts, as produced by the watchdog tools. Looking at these logs, the correct behaviour of the central uploader machine can be controlled, so that an alarm system, based on that information, can recognize if some exportations were not successful and, eventually, inform the end-user of the error occurred. A screenshot of the page is shown in Figure~\ref{fig:WatchDog}.    
\end{enumerate}

Figure \ref{fig:PopConAct} reports all the transactions towards the condition database accounts that occurring in a month of cosmic data taking. 
As the summary plot points out, almost all sub-detectors used PopCon to upload calibration constants to the condition databases.
An average of one hundred PopCon applications per day were run during the test runs in Summer/Fall 2008, hence one hundred connections per day to the condition databases took place.

During the entire year 2008, the total amount of condition data written in the production database was approximatively 1 TB. 
Indeed, no network problems, neither for the online-offline streaming, nor for Frontier were detected. 
All the conditions and calibrations were properly evaluated during the cosmic ray test-runs in 2008, leading to several global tags that were used for the reconstruction and the analysis of the cosmic ray data by the whole CMS community.

\section{Conclusion}

A robust system was set-up in order to upload, store and retrieve all calibration constants for the CMS experiment. 
The system relies on ORACLE databases for data storage, and on the POOL-ORA technology, embedded in the PopCon farmework (written in \texttt{C++}) integrated in the overall CMSSW architecture, for data handling. 
The whole chain was deployed and tested successfully during 2008 commissioning exercises with cosmic rays: these tests have demonstrated that the system we described is stable and robust enough for the 2009-2010 collision data taking.

\end{document}